# Columba-Hypatia: Astronomy for Peace

Francesca FRAGKOUDI[*1], Marja SEIDEL[*2], GalileoMobile[*3]
and the Association for Historical Dialogue and Research[*4]

**Abstract.** "Columba-Hypatia: Astronomy for Peace" is a joint astronomy outreach project by GalileoMobile and the Association for Historical Dialogue and Research (AHDR) which takes place on the island of Cyprus. The project aims to inspire young people, through astronomy, to be curious about science and the cosmos, while also using astronomy as a tool for promoting meaningful communication and a culture of peace and non-violence. We conduct educational astronomy activities and explore the cosmos with children and the public, bringing together individuals from the various communities of Cyprus 'under the same sky' to look beyond borders and inspire a sense of global citizenship.

## 1. Introduction

The "Columba-Hypatia project: Astronomy for Peace"[1] is a joint astronomy outreach project by GalileoMobile and the Association for Historical Dialogue and Research (AHDR) taking place on the island of Cyprus. The goal of the project is to use astronomy as tool to promote a culture of peace and non-violence on the island. The two main ethnic communities of the island, the Greek-Cypriots (GC) and Turkish-Cypriots (TC) have been living separated from one another for over 50 years in a post-conflict environment. While border crossings were opened in 2003 — which has allowed access to both sides of the island — interaction and cooperation between the two communities is still rare.

Through this grass-roots initiative, the Columba-Hypatia project aims to inspire children and the general public to be more curious about science and the Universe, while also bringing together children and adults from the two communities to break down prejudices and misconceptions, thus promoting meaningful communication.

The project had a pilot run in 2016 and subsequently ran throughout the year of 2017. It was funded by the International Astronomical Union's Office of Astronomy for Development (OAD). Additional educational materials were donated by the European Southern Observatory's education and Public Outreach Department (ESO ePOD), Universe Awareness (UNAWE) and telescopes were donated by Meade instruments.

*1 Max-Planck-Institut für Astrophysik,
ffrag@mpa-garching.mpg.de
*2 The Observatories of the Carnegie Institution for Science
*3 www.galileo-mobile.org; info@galileomobile.com
*4 www.ahdr.info; ahdr@ahdr.info

## 2. GalileoMobile & AHDR

GalileoMobile is an itinerant, non-profit outreach project that shares astronomy with students and teachers in schools and communities worldwide. The team is comprised by a group of volunteer astronomers, educators and science communicators around the world. Since its inception in 2008, GalileoMobile has reached 1,400 teachers and 16,000 students, donating more than 100 telescopes and organising public events for more than 2,500 people in 14 countries. GalileoMobile shares astronomy across the world in a spirit of inclusion, sustainability, and cultural exchange to create a feeling of unity under the same sky.

AHDR is an intercommunal organisation based in Cyprus whose mission is to contribute to the advancement of historical understanding amongst the public and more specifically amongst children, youth and educators by providing access to learning opportunities for individuals of every ability and every ethnic, religious, cultural and social background, based on the respect for diversity and the dialogue of ideas.

## 3. Execution of Project

The main phase of the project began in 2017, and ran throughout the entire year. The project involved mono-communal school visits — where trainers visited GC and TC schools separately in Cyprus — and bi-communal activity days, were children from GC and TC schools came together in the "buffer zone" at the Home for Cooperation, to meet each other and participate in astronomy activities. The buffer zone is a United Nations controlled area between the two communities. We also carried out afternoon/evening events for youth as well as for the general public, such as a bi-communal summer camp, astronomy outreach talks, astroparties, etc. The main focus of the project was on the mono-communal school visits and bi-communal activities, since these could reach the most diverse audience and have the largest impact. The numbers of participants are listed in Table 1. For the astronomy activities we used the GalileoMobile Handbook of activities[2] adapted for the activities and translated into Greek and Turkish.

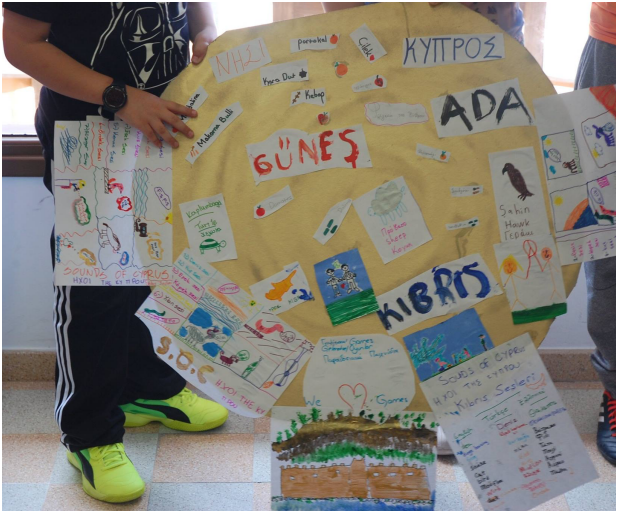

Fig. 1. The Cyprus Golden Record, constructed by the participating children in one of the bi-communal activity days. *Credit: Columba-Hypatia project.*

### 3.1. Mono-communal visits

During the mono-communal visits we focused on introducing the children to the project, and to basic astronomical concepts. The activities were chosen to give the children an idea of the place of the Earth in the context of modern astronomy, to introduce them to the vast scales and sizes of the Universe (e.g. comparing the sizes of stars and planets, while also observing the Sun through solar telescopes). We also carried out activities with the UNAWE Earth Ball to show how the Earth looks from space and introduce the concept of man-made borders.

### 3.2. Bi-communal days

During the bi-communal days, children from the GC and TC communities came together in the buffer zone to participate in astronomy activities. The main two activities carried out were "Building a Cyprus Golden Record" and "Building Constellations in 3D". The former, allowed the children to be split into mixed groups and discuss between them what they would send to an alien civilisation as a representation of the whole of the island of Cyprus (see Fig. 1). This allowed the children to get to know each others' cultures, and to discover the many similarities between them.

The latter activity introduced the concept of viewing objects (in this case, a constellation), or indeed situations, from different angles, and how our perspective can change depending on our viewpoint.

Table 1. Number of participants

|  | Number of participants |
|---|---|
| **School children** | ~190 |
| **Teachers** | ~20 |
| **Youth** | ~100 |
| **General Public** | ~150 |
| **Documentary views** | ~4300 |

### 4. Impressions, Feedback & Evaluation

The project was well received both by the teachers as well as the children who participated. The participants showed enthusiasm and genuine curiosity about members of the other community, and reported feelings of improved understanding and empathy of each-other after the bi-communal visits (c.f. the short documentary of the project [3]). In order to measure this effect in future instalments of the project we plan to evaluate the impact of the project and the "Pale blue dot effect". This latter effect is one in which "*…knowing one's place in the Universe alters perception and induces more empathy towards fellow humans.*"[4]

### 5. Summary

Through the Columba-Hypatia project, we experienced how effective astronomy is in promoting a feeling of global citizenship and a culture of peace, enabling children to broaden their views and interactively explore together their place on Earth and, specifically for our project, on the island of Cyprus. The project was implemented successfully during 2017, and current and future instalments are ongoing and planned for the coming years.